\documentclass{aa}  
\usepackage{graphicx}
\usepackage{txfonts}
\usepackage{natbib}
\bibpunct{(}{)}{;}{a}{}{,} 
\begin{document}
   \title{Resolved debris disk emission around $\eta$ Tel: a young Solar System
   or ongoing planet formation?}

   \author{R. Smith
          \inst{1}
          \and
          L.J. Churcher\inst{1} \and M.C. Wyatt\inst{1} 
          \and M.M. Moerchen \inst{2,3} \and C.M. Telesco \inst{2}           
          }

   \offprints{R. Smith}

   \institute{Institute of Astronomy (IoA), University of Cambridge,
              Madingley Road, Cambridge, CB3 0HA, UK\\
              \email{rsed@ast.cam.ac.uk}
         \and
             Department of Astronomy, University of Florida, 
             211 Bryant Space Science Center,  P.O. Box 112055, 
             Gainesville, FL, 32611-2055, USA\\
         \and
             European Southern Observatory, 
             Alonso de Cordova 3107, Casilla 19001, Vitacura, 
             Santiago 19, Chile\\
             }

   \date{Draft 28 October 2008}

 
  \abstract
   {}
     {60\% of the A star members of the 12 Myr old $\beta$ Pictoris moving group
     (BPMG) show significant excess emission in the mid-infrared, several
     million years after the proto-planetary disk is thought to disperse.
     Theoretical models suggest this peak may coincide with the formation of 
     Pluto-sized planetesimals in the disk, stirring smaller bodies into
     collisional destruction.
     Here we present resolved mid-infrared imaging of the disk of $\eta$ Tel
     (A0V in the BPMG) and consider its implications for the state of planet 
     formation in this system.}
     {$\eta$ Tel was observed at 11.7 and 18.3$\mu$m using T-ReCS on
       Gemini South. The resulting images were compared to simple disk
       models to constrain the radial distribution of the emitting material.}
     {The emission observed at 18.3$\mu$m is shown to be significantly
     extended beyond the PSF along a position angle 
     $8^\circ$. This is the first time dust emission has been
       resolved around $\eta$ Tel. 
     Modelling indicates that the extension arises from an edge-on disk of
     radius 0.5 arcsec ($\sim$ 24 AU).
     Combining the spatial constraints from the imaging with those from
     the spectral energy distribution shows that $>$50\% of the 18$\mu$m
     emission comes from an unresolved dust component at $\sim$4 AU.}
     {The radial structure of the $\eta$ Tel debris disk is 
     reminiscent of the Solar System, suggesting that this is a young
     Solar System analogue.
     For an age of 12Myr, both the radius and dust level of the extended
     cooler component are consistent with self-stirring models
     for a protoplanetary disk of 0.7 times minimum mass solar nebula.
     The origin of the hot dust component may arise in an asteroid belt
     undergoing collisional destruction, or in massive collisions in
     ongoing terrestrial planet formation.}


   \maketitle
%

\section{Introduction}

Debris disks around nearby stars are often discovered through the
detection of emission beyond that predicted to come from the stellar
photosphere in the infrared or sub-millimeter \citep[see][for a recent
  review]{wyattreview}. 
The detected emission arises from the absorption and subsequent re-emission of 
starlight at longer wavelengths by dust particles in the disks.
These dust grains are believed to be continuously reproduced by the collisional
destruction and/or sublimation of larger bodies (planetesimals), as
the lifetimes of such small grains due to removal by radiation
pressure or Poynting-Robertson (P-R) drag are small compared to the
age of the star \citep{backman}.
Studying these debris disks can give clues about how 
the circumstellar region has evolved during the star's formation and
subsequent lifetime, how any planetary system around the star may have
formed and evolved, and the current status of the system \citep[see 
e.g.,][]{wyatt99}. 

$\eta$ Tel (HD\,181296) is an A0Vn member of the $\beta$ Pictoris moving
group, with an age estimated to be 12$^{+8}_{-4}$ Myr
\citep[age estimated for moving group in][]{zuckermanEA}. 
It has an M7/8V companion at an offset of 4\arcsec at a position angle of
160$^\circ$ to the star \citep{lowrance}.
Hipparcos listed parallax measurements put this system at a distance
of 48pc, giving a projected offset of 192 AU for the companion. 
\citet{backman} first identified this star as a debris disk candidate
based on IRAS  measurements which indicates that excess emission is
present at 12, 25 and 60$\mu$m.
The excess measurement at 12$\mu$m is particularly rare \citep{aumannprobst},
as debris disks are more typically characterised by dust at several 10s of AU
which is too cool to emit at 12$\mu$m. Indeed the majority of
  A star sources with disk emission resolved at thermal wavelengths
  have disks extended on 100s of AU scales \citep[$\beta$ Pic,
    HR4796A, 49 Ceti, HD\,141596 and 
    HD\,32297][]{backman,jayawardhana,wahhaj,fisher,moerchenb}.
Nevertheless, examples of A stars with dust confirmed to lie within
10AU are known (e.g., $\zeta$ Lep, \citealt{moerchen}), and there
remains debate as 
to whether such hot emission can originate in the steady-state destruction
of planetesimal belts \citep{wyattsmith06} or in collisions between
growing protoplanets that are expected during terrestrial planet
formation \citep{kenyon02}. \citet{chen06} presented IRS observations
of $\eta$ Tel showing that the excess emission cannot be fit by a
single temperature of dust as the spectrum is too flat.  
They interpret this as evidence for two
temperatures of dust, but from the emission spectrum alone it is not
possible to tell if this arose from dust in two separate radial
locations around the star, or alternatively two different grain
populations at the same radial location.  The aim of this paper is to
use resolved imaging of the disk to allow a differentiation between
these alternatives.  In addition these images will confirm
  that the excess emission is in fact centred on the primary star and
not the binary, although due to the brightness of the 12$\mu$m flux it
is unlikely that the emission arises from dust around the M7/8
companion. 

The age of this system and membership of the moving group make it very
interesting in terms of evolutionary studies of debris disks. 
Recent work by \citet{Currie} has shown that excess emission at 24$\mu$m
around A stars increases from 5--10 Myr, peaks around 10--15 Myr and
then declines with age. The cause of the peak is not yet known \citep[see review
  in][]{wyattreview}, but is suggested to be the delayed formation of
Pluto-sized bodies further from the star, as these bodies are
necessary to stir the disk so that collisions are destructive thus
releasing large quantities of dust \citep{kenyon04II}.
The 12 Myr $\beta$ Pictoris moving group has 5 A star members, 3 of which are 
known to have debris disks that exhibit excess emission in the mid-infrared
($\beta$ Pictoris, HD\,172555 and $\eta$ Tel; \citealt{chen06, rebull}), 
and the proximity of this group means it offers the chance
to assess the origin and diversity of the peak.  

Here we present 11.7 and 18.3$\mu$m imaging of $\eta$ Tel
taken with the T-ReCS instrument on Gemini South which shows a resolved
debris disk at 18.3$\mu$m.
We confront these observations with models to determine the optimal disk 
parameters and discuss the implications of the inferred structure for the
status of planet formation in this system, both by comparison with
the self-stirred models of \citet{kenyon04II} and with our understanding of
the early Solar System. 


\section{Observations}

The source was observed under proposal GS-2007A-Q-45 using T-ReCS
on Gemini South with 
filter Qa ($\lambda_c$ = 18.3 $\mu$m, $\Delta\lambda$ = 1.51 $\mu$m) and
Si-5 ($\lambda_c$  = 11.66 $\mu$m, $\Delta\lambda$ = 2.13 $\mu$m),
which we refer to as Qa and Si-5 observations henceforth. The
pixel scale of the T-ReCS imager is 0\farcs09 with a total field of
view of 29\arcsec $\times$ 22\arcsec. The observations were taken in
parallel chop-nod mode with a chop throw of 10\arcsec and chop
position angle of 55$^\circ$ (East of North) at Q, 100$^\circ$ at N\@.
The Qa observations were taken over two consecutive 
nights (1$^{\rm{st}}$ and 2$^{\rm{nd}}$ July 2007), with the Si-5 
observations taken on one night only (12$^{\rm{th}}$ July 2007).
Total integration time was 9120s 
(4560s on source) at Qa, and 912s (456s on source) at Si-5\@. 
Observations of standard stars HD\,196171 (spectral type K0III)
and HD\,179886 (spectral type K3III), both listed in \citet{cohen},
were made to calibrate the photometry and also determine the PSF of the 
observations.  The order of observations 
and integration times are shown in Table \ref{tab:basic_obs}. 

These data for $\eta$ Tel were taken in multiples of 304s-long
integrations, each consisting of 7 complete nod cycles (ABBA).  The data
were reduced using custom routines described in \citet{smithhot}.  Data
reduction involved determination of a gain map
using the mean values of each frame to determine pixel responsivity
(masking off pixels on which emission from the source could fall,
equivalent to a sky flat).  In
addition a dc-offset was determined by calculating the mean pixel
values in columns and rows (excluding pixels on which source emission
was detected) and this was subtracted from the final image to ensure a flat
background. Pixels showing high or low gain, or those
which showed great variation throughout the observation were masked
off. In order
to avoid errors in co-adding the data which could arise from
incorrect alignment of the images, we used re-binned images in which
each pixel was a fifth the size of the T-ReCS image pixels to determine
an accurate centre of the stellar images, aligned and co-added these
images and then re-binned to the original pixel size.  The re-binning
was performed using bilinear interpolation across the array. This
technique was used both to determine a final image of $\eta$ Tel and
standard star for each band.  To ensure that background noise
  did not affect the re-binning process the same procedure was applied
with an additional step of smoothing the images with a Gaussian kernel
of FWHM equal to that of the standard star image before re-binning.
It was found that the relative sizes of the final images of $\eta$ Tel
and the standard stars and ellipticity of the images was unchanged by
this additional step, indicating that the re-binning was not affected
by background noise.  This is as to be expected in  data such as this
where the detector super-samples the PSF. 

\begin{table}[!h]
\centering
\begin{tabular}{*{5}{|c}|} \hline Date & Object & Filter & Integration
& Calibrated  \\  &  &  & time (s) & flux, mJy  \\ \hline 
01/07/2007 & HD\,196171 & Qa & 152 & 6058 \\
01/07/2007 & $\eta$ Tel & Qa & 1824 & 350 $\pm$ 11 \\
01/07/2007 & HD\,196171 & Qa & 152 & 6058 \\
01/07/2007 & $\eta$ Tel & Qa & 1824 & 347 $\pm$ 11 \\
01/07/2007 & HD\,196171 & Qa & 152 & 6058 \\ \hline
02/07/2007 & HD\,196171 & Qa & 152 & 6058 \\
02/07/2007 & $\eta$ Tel & Qa & 1824 & 345 $\pm$ 10 \\
02/07/2007 & HD\,196171 & Qa & 152 & 6058 \\
02/07/2007 & $\eta$ Tel & Qa & 1824 & 350 $\pm$ 11 \\
02/07/2007 & HD\,196171 & Qa & 152 & 6058 \\
02/07/2007 & $\eta$ Tel & Qa & 1824 & 331 $\pm$ 10 \\ 
02/07/2007 & HD\,196171 & Qa & 152 & 6058 \\ \hline
12/07/2007 & HD\,179886 & Si-5 & 304 & 4417 \\ 
12/07/2007 & $\eta$ Tel & Si-5 & 912 & 342 $\pm$ 18  \\
12/07/2007 & HD\,179886 & Si-5 & 304 & 4417 \\ \hline 
\end{tabular}
\caption[The Observations of $\eta$ Tel]{\label{tab:basic_obs} Observations 
  in order and total
  integration times of the observations taken under proposal
  GS-2007A-Q-45.  Note that on-source integration time is half the
  total integration time listed in the table.}
\end{table}

\section{Results} 

\begin{figure*}
\begin{minipage}{1cm}
Qa
\end{minipage}
\begin{minipage}{6cm}
\center{$\eta$ Tel}
\includegraphics[width=6cm]{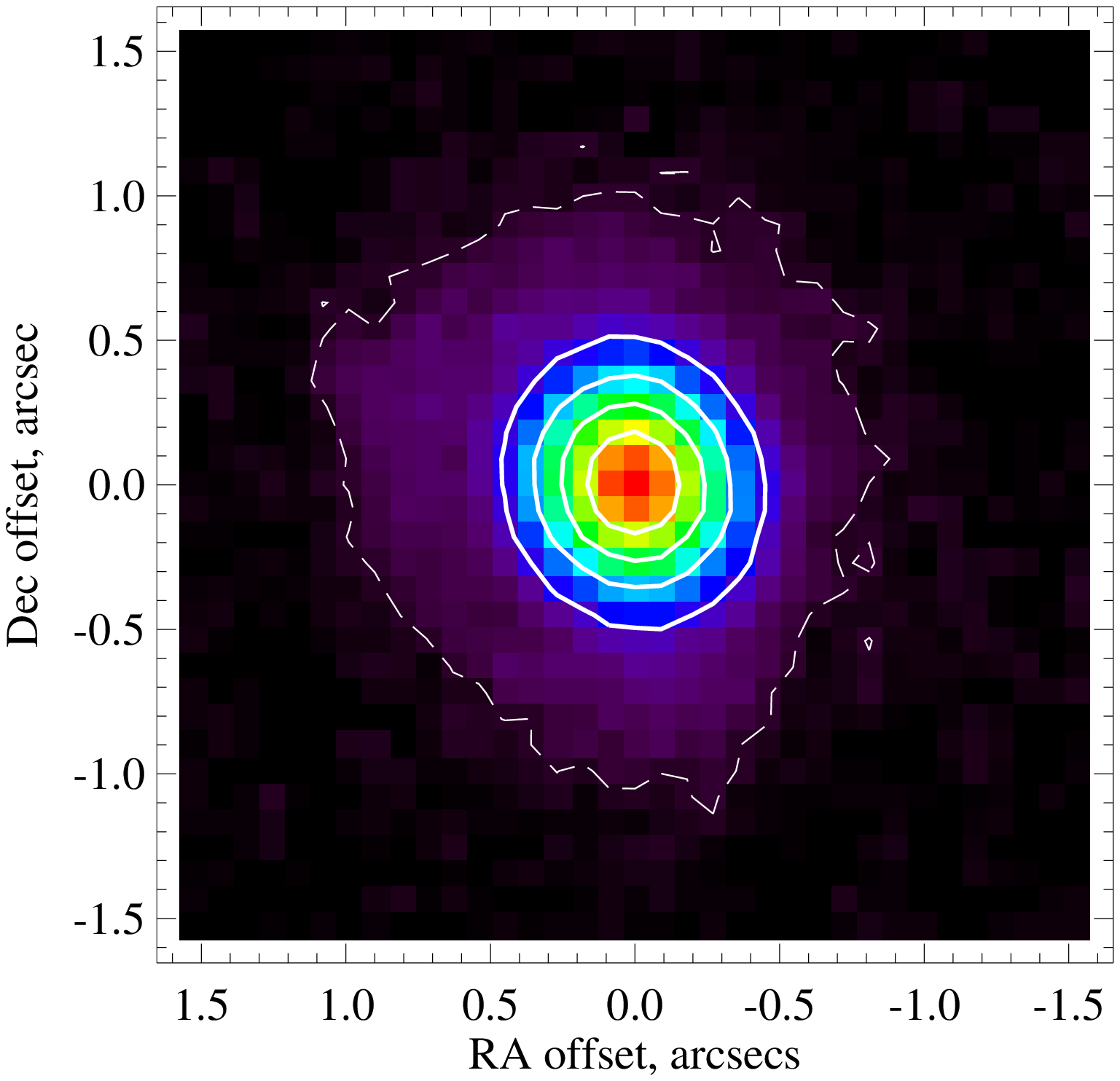}
\end{minipage}
\hspace{-0.5cm}
\begin{minipage}{6cm}
\center{Standard}
\includegraphics[width=6cm]{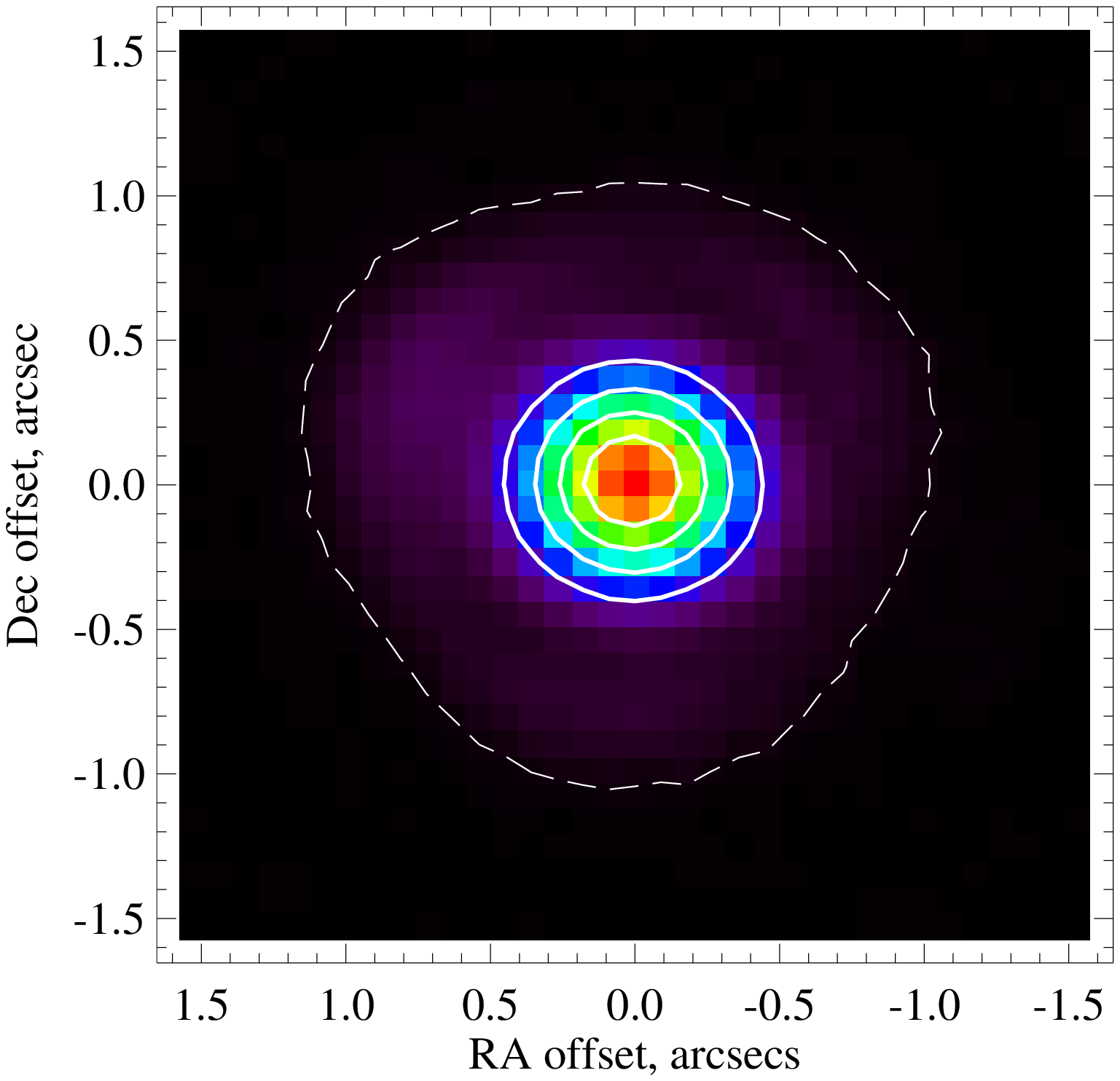}
\end{minipage} 
\hspace{-0.5cm}
\begin{minipage}{6cm}
\center{Residual}
\includegraphics[width=6cm]{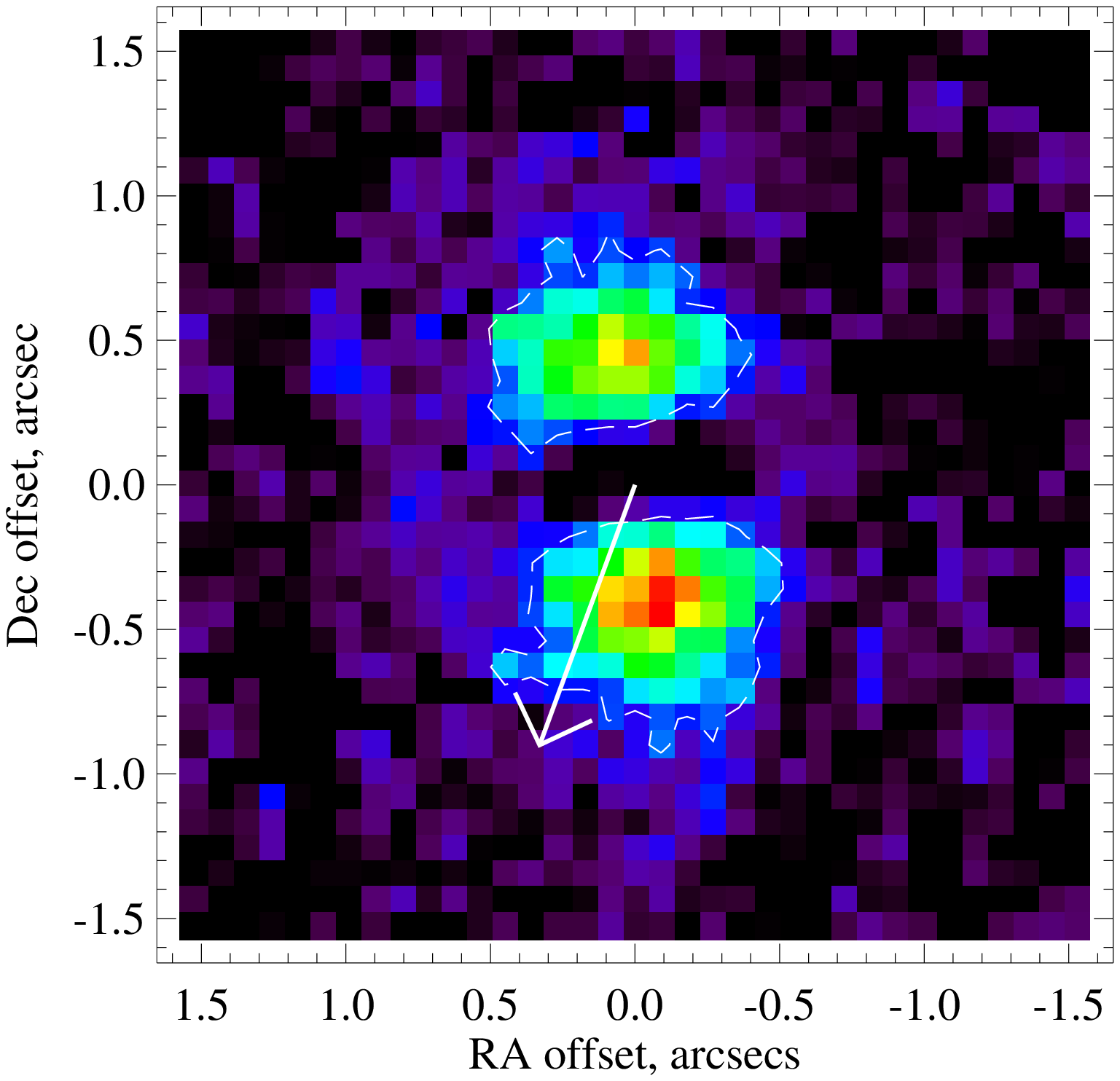}
\end{minipage} \\
\begin{minipage}{1cm}
Si-5
\end{minipage}
\begin{minipage}{6cm}
\includegraphics[width=6cm]{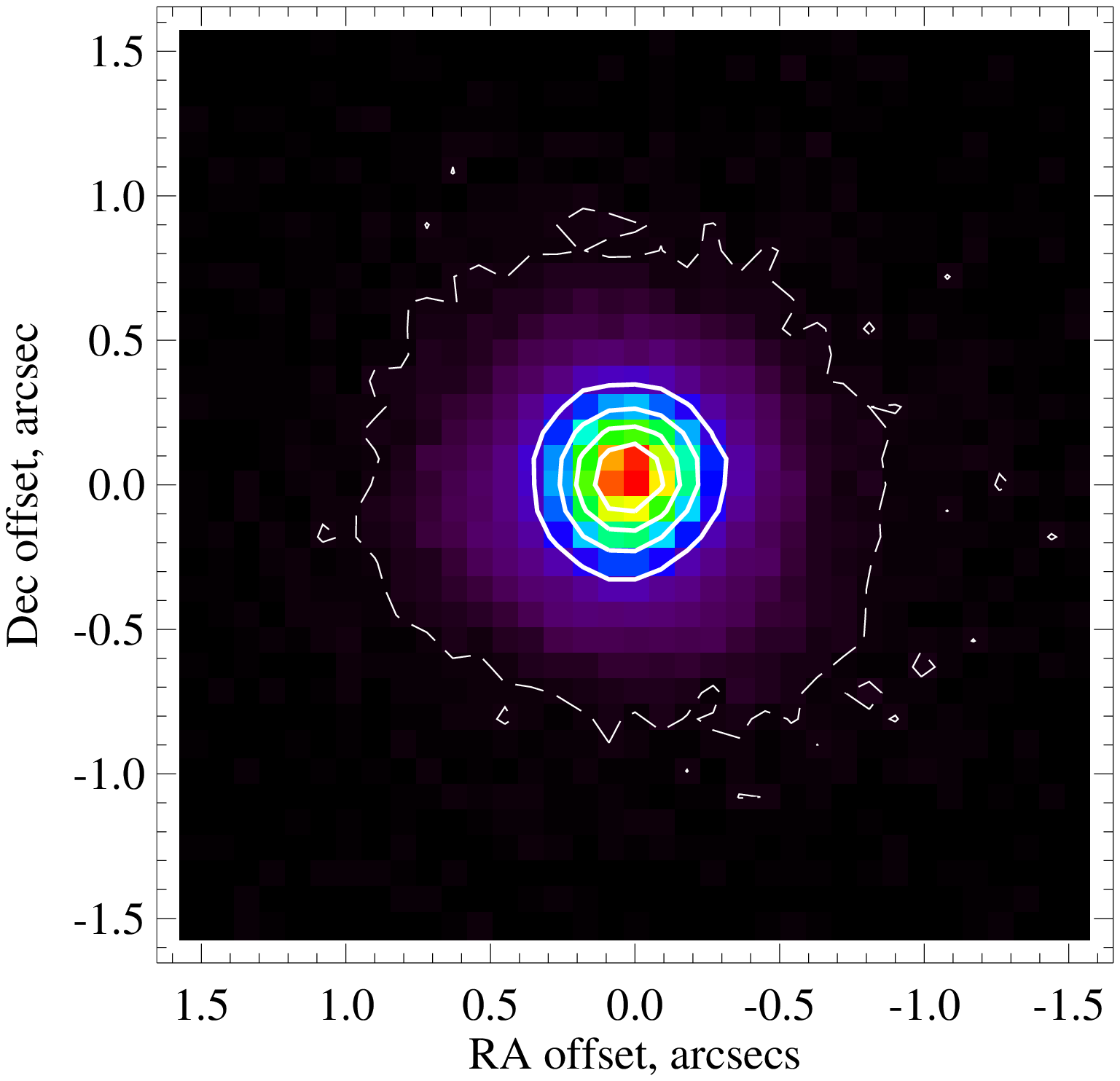}
\end{minipage}
\hspace{-0.5cm}
\begin{minipage}{6cm}
\includegraphics[width=6cm]{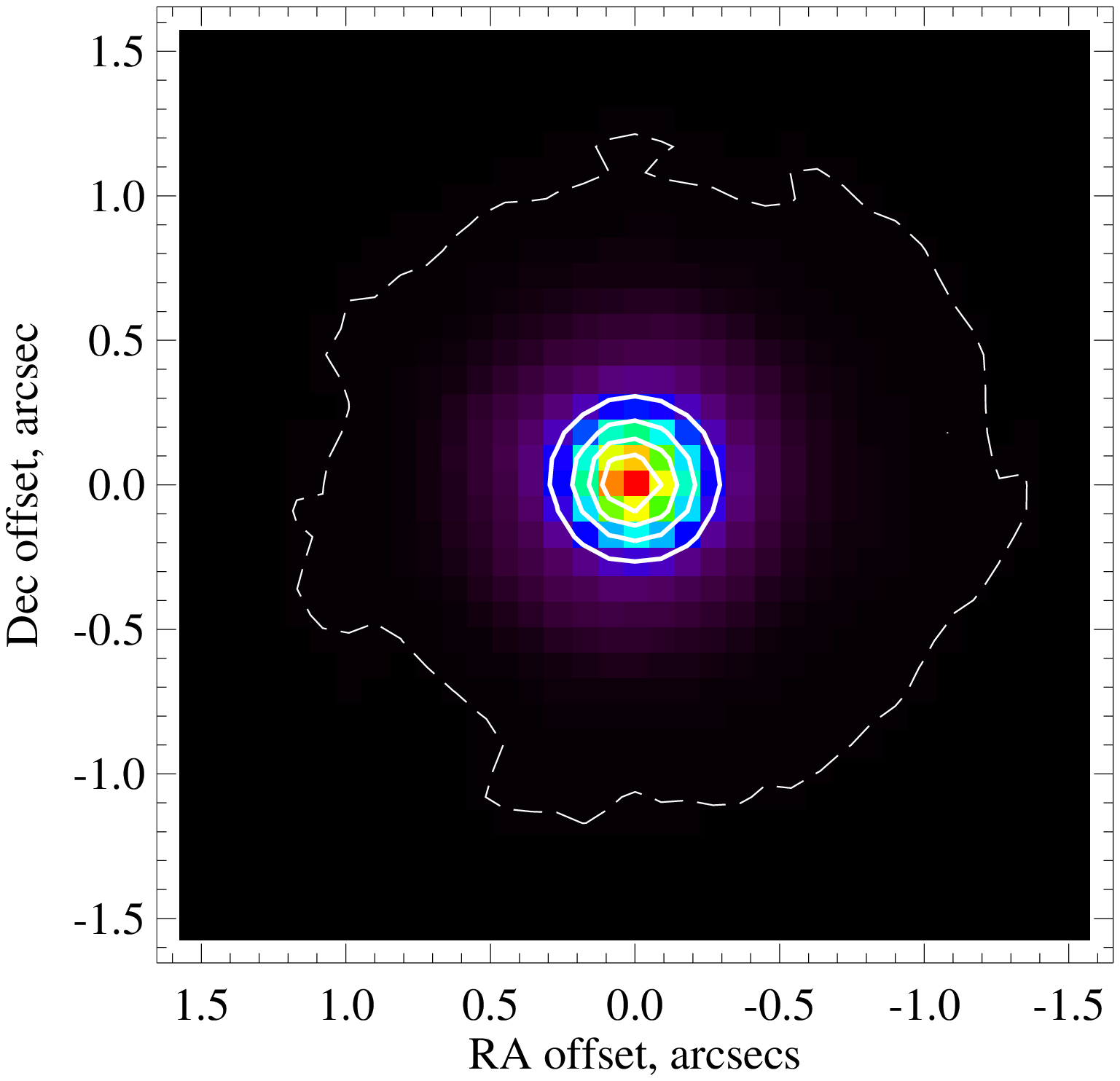}
\end{minipage}
\hspace{-0.5cm}
\begin{minipage}{6cm}
\includegraphics[width=6cm]{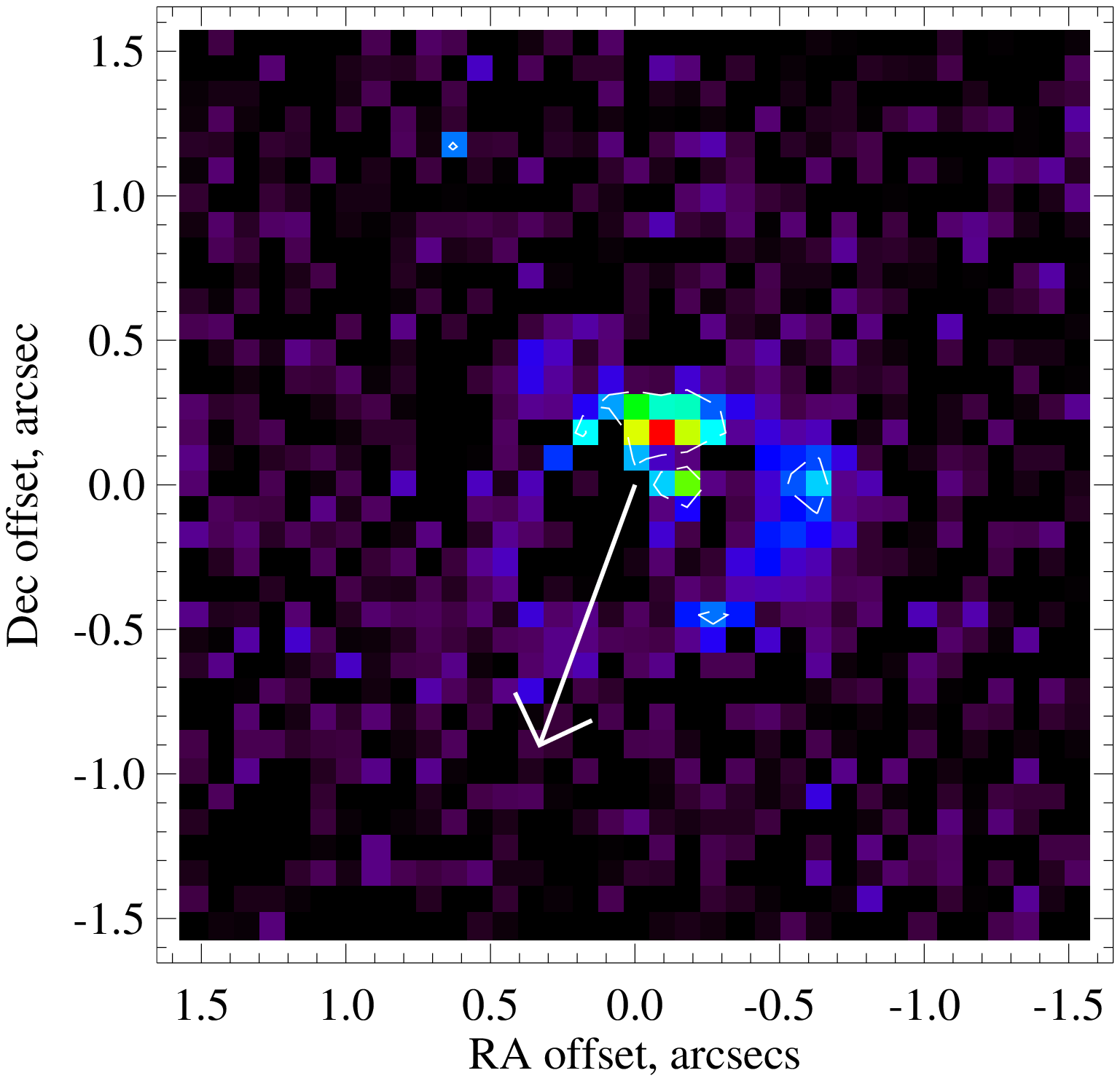}
\end{minipage}
\vspace{1cm}
\caption[The Final T-ReCS Images of $\eta$ Tel]{\label{fig:cons} The co-added 
  final images from the Qa observations (top row) and the Si-5
  observations (bottom).  Orientation of the images is 
  North up, East left. Left: The total co-added images
  of $\eta$ Tel. Middle: The co-added images of the standard star forming
  our reference PSF\@.  Note the distinct ellipticity with major axis in
  the North-South direction seen in the Qa image of $\eta$ Tel is
  not seen in Si-5 or in the standard star images. Solid line contours are at 
  the 20\%, 40\%, 60\%, and 80\% of the peak levels for all
  images. Peaks are at a level of 650 mJy/arcsec$^2$ for $\eta$ Tel at
  Qa, 1220 mJy/arcsec$^2$ for $\eta$ Tel at Si-5, 13990 mJy/arcsec$^2$ for
  the standard star at Qa and 20670 mJy/arcsec$^2$ for the standard at Si-5.  
  Right: The residual emission after subtraction of
  the total standard star image scaled to the peak of the image of $\eta$ Tel.
  The dashed contour indicates the 3$\sigma$ per pixel level
  (from background noise). 
  The colour scale is linear from 0 to the maximum pixel value in each
  image. The
  residual images have peak brightness of 88 mJy/arcsec$^2$ at Qa and
  35 mJy/arcsec$^2$ at Si-5. The
  arrow indicates the direction of the M7/8V binary which lies at 4\arcsec,
  PA$^\circ160$. Note that the emission seen in the Si-5 residual image is
  not significant ($< 1 \sigma$ significance) when the variation in the PSF
  is considered.}
\end{figure*}

The final co-added images of $\eta$ Tel in both filters and the
standard star for PSF reference are shown in Figure \ref{fig:cons}.
The contours shown in the images are at 20\%, 40\%, 60\% and 80\% of
the peak in each case (peak value in Qa 5.24mJy/pixel, in Si-5
9.89mJy/pixel).  
Also shown in the right-hand image for each
row is the residual emission around $\eta$ Tel observed when
subtracting the standard star image as a PSF reference scaled to the
peak of the $\eta$ Tel image.  The contours and residuals are
discussed further below. 

Photometry was performed using 1\farcs0 radius circular apertures centred on
the stellar image. The centre of the image was determined through a
2-dimensional Gaussian fit. An average of the calibration factors
determined from the standard star observations  taken on the night was
used for corresponding photometric calibration. Statistical noise was
determined using an annulus with inner radius 1\arcsec and outer
radius 2\arcsec centred on the source, resulting in uncertainties of
0.13 and 0.06 mJy/pixel at Qa and Si-5 respectively. The calibration factors
were found to vary by 4\% and 2\% on the first and second nights of Qa
band observations, respectively, which is consistent with the
variation of 3\% found between fluxes measured on individual 304s
integrations on $\eta$ Tel (which is bright enough for an $\sim$60 sigma
detection even in such a short integration). 
Although the calibration factors varied by just 5\% for the two observations
of standards in the Si-5 filter, analysis of the individual 304s integrations on
$\eta$ Tel (each of which resulted in a 78 sigma detection) showed that
photometric accuracy was in fact at a level of 13\%. Our photometry
yields total fluxes of 342 $\pm$ 44 mJy in Si-5 (with S/N of 292) and 345
$\pm$ 15 mJy in Qa (with S/N of 129), including both calibration and
statistical noise, where the S/N given in brackets indicates the level
of statistical noise.  Standard stars were observed at a
  similar airmass to $\eta$ Tel.  In addition, calibration levels were
  compared to airmass for the standard star observations and no
  correlation was found, therefore no correction for extinction was
  applied to calibration.  
Using a Kurucz model profile scaled to the 2MASS K band magnitude,
the stellar flux in filters Si-5 and Qa is expected to be 282 mJy and
114 mJy, respectively. 
Thus the photometry gives an excess of $60 \pm 44$mJy and $231 \pm 15$mJy
in Si-5 and Qa.
The IRS spectrum of \citet{chen06} after subtraction of the above
mentioned photospheric emission gives excess emission of 150$\pm$5 and
257$\pm$2 mJy, respectively, at these two wavelengths and so the
photometry presented here is in agreement within the T-ReCS photometric
errors with the IRS results at the 2$\sigma$ level.  
Thus it does not appear that the IRS spectrum includes any emission centred
on the star outside the 1\arcsec aperture used here for photometry, as might
happen in the larger IRS beam (extracted along a 3\farcs7 slit).

Figure \ref{fig:cons} shows the final co-added images of $\eta$ Tel and
the standard star for a PSF reference.
The top line of this figure shows the Qa total image of $\eta$ Tel and
the standard.
An elliptical shape can clearly be discerned, and a 2-dimensional Gaussian fit
to the image has an ellipticity of 0.117 $\pm$ 0.007 with the major
axis at $19.9\pm2.0^\circ$ East of North;
the same ellipticity was seen on the two separate nights of Qa observing.
For comparison, the ellipticity of the final standard image is 0.084 $\pm$ 0.005
with the major axis at 83$\pm2^\circ$.  This slightly elliptical shape
of the PSF is seen across the standard star integrations (see later
this section) and does not coincide with the extension seen in the
$\eta$ Tel images. 
As all Qa observations were performed with no on-sky rotation and
chop-nod performed at 55$^\circ$, the direction of the
extension is unlikely to be the result of chop smearing.

\begin{figure*}
\begin{minipage}{8cm}
\includegraphics[width=8cm]{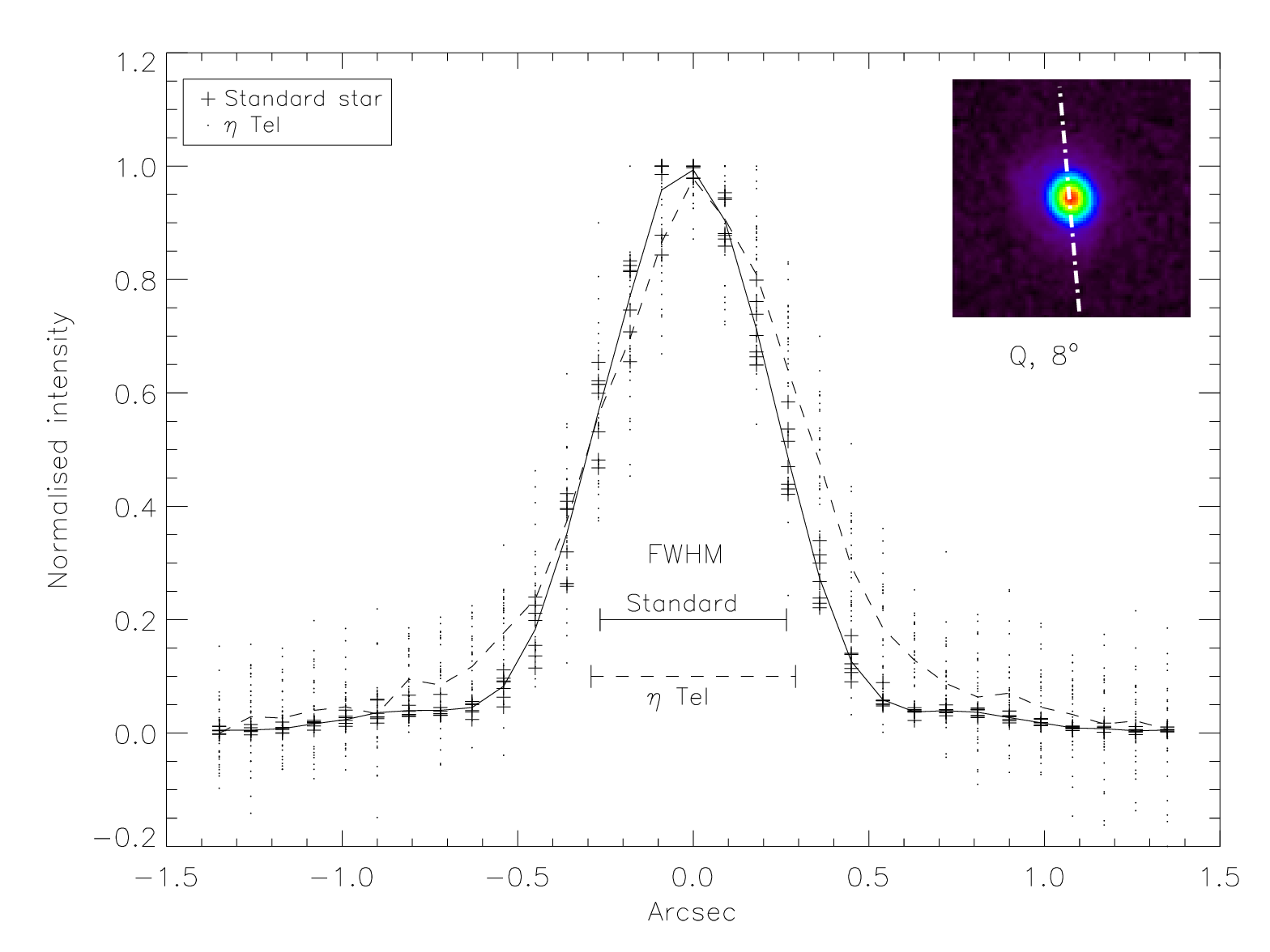}
\end{minipage}
\hspace{1cm}
\begin{minipage}{8cm}
\includegraphics[width=8cm]{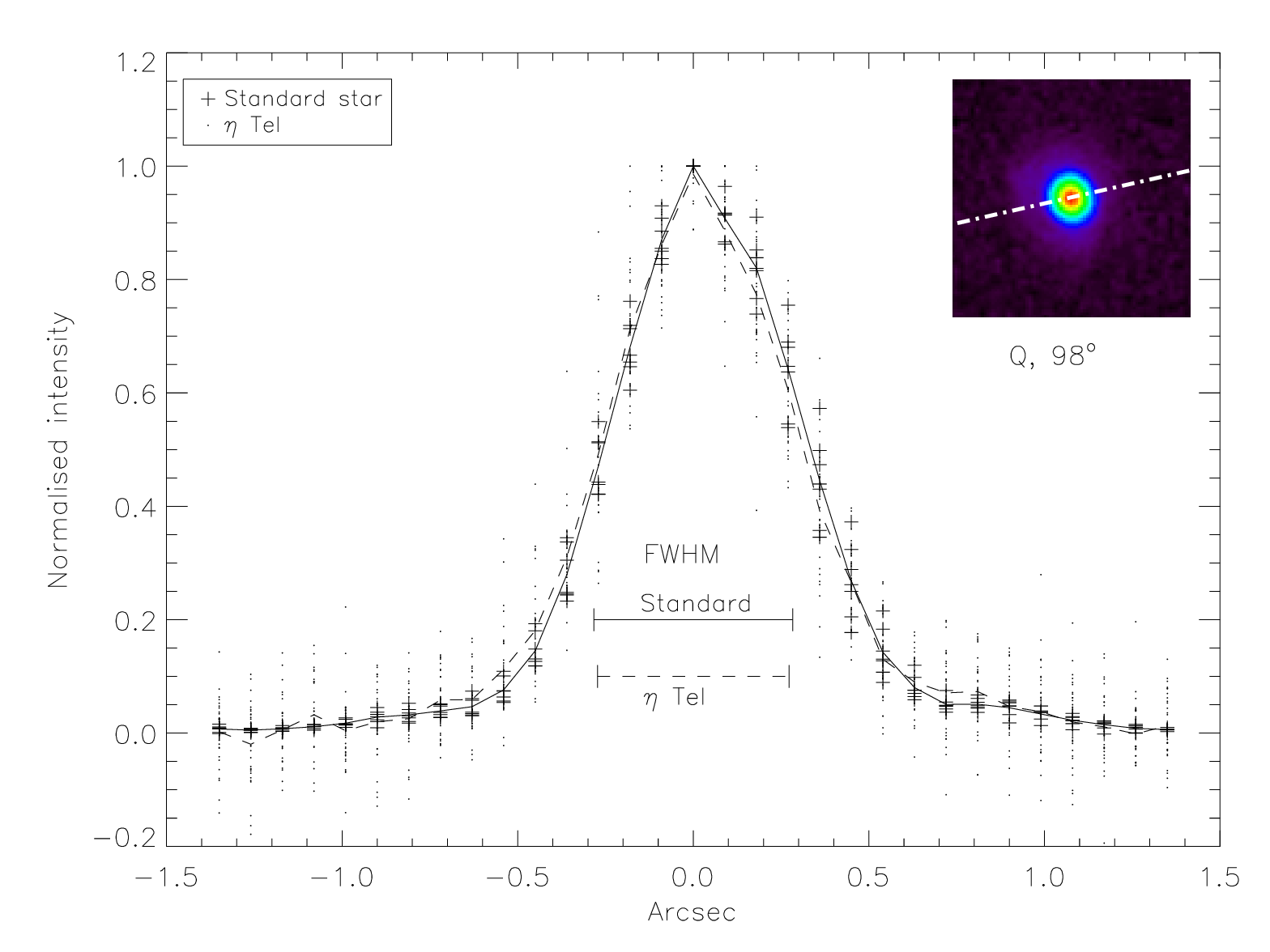}
\end{minipage}
\caption[Gaussian Fits to the Line Cuts Through the Images of $\eta$ Tel 
  and Associated Standard Stars]{\label{fig:profiles} The 
  profiles of line cuts through the total images, at 8$^\circ$ (left)
  and 98$^\circ$ (right) at Qa.  The median FWHM measurements taken
  over all sub-integrations from Gaussian fits to the profiles at the
  two angles are indicated for reference.  Also shown are averaged
  profiles from all the line cuts for both $\eta$ Tel and the standard
  star. Observe that the profile for $\eta$ Tel at 8$^\circ$ is much
  broader in the wings than the standard star. The 1$\sigma$ error bar
  shows the size of $\pm1\sigma$ background error per pixel from the $\eta$ Tel
  image.  Note that multiple intensity profiles for the science target and
  standard are shown here.  These are taken
  from longer integrations than noted in the text to reduce the number
of profiles and improve the clarity of the figure. }
\end{figure*}

Also shown in Figure \ref{fig:cons} is the residual emission of $\eta$
Tel following a subtraction of the PSF reference standard star image
scaled to the peak of the $\eta$ Tel image, which subtracts a total of
252mJy from the image.
Although this subtraction method is likely to remove some of the disk
emission (some of which contributes to the peak surface brightness),
it does highlight the location of the non-pointlike emission.
Peaks are seen in the Qa residual image that are centred at
0\farcs45 (21.6AU), 5.2$^\circ$ East of North,
and 0\farcs40 (19.2AU), 190.5$^\circ$ East of North.
Symmetrical peaks in the residual emission either side of the star are
exactly what would be expected for an edge-on ring
\citep[][although also note that we cannot rule out a less inclined
  ring without detailed modelling of the disk, see section
  4.]{telesco00}. Since the residual peaks are more representative of
the extension seen than the 2-dimensional Gaussian fit discussed
above, hereafter the extension is considered to be at a position angle
of $\sim$8$^\circ$. Thus the extension is at $\sim$28$^\circ$ with
respect to the direction of the binary  (shown with an arrow),
suggesting if this arises from an edge-on disk, then that disk is not
in the same plane as the binary. The flux in the residual peaks totals
80 $\pm$ 5 mJy (measured in 0\farcs4 radius apertures; the
  error term includes background and calibration uncertainty added in
  quadrature), which as expected is below that of the excess emission
231$\pm$15mJy measured in our photometry.  An estimate of the
  level of disk emission which may have been removed from the
  residuals image by our method of PSF subtraction can be determined
  from consideration of how much the two peaks may have contributed to
  the central pixel.  Using the PSF model scaled to the residual peaks
  this was calculated as 0.3mJy (as compared to the total peak before
  subtraction of 5.24mJy). This method suggests the extended emission
  represents a minimum of 6\% of the peak emission, as if the extended
  emission is truly a disk lying across the star, this method will
  again underestimate the disk flux in the central pixel). 
  A more detailed model of the
extended emission seen in the Qa image is discussed in section 4. 

To assess the temporal variability of this FWHM and the significance
of the extensions, as well as to search for a
dependence on time or airmass, linecuts were performed at position
angles of 8$^\circ$ and 98$^\circ$ for individual 
nod set images (22s) for the standard star and two nod set images
(43s) for $\eta$ Tel. 
Two methods of fitting the cuts, by a Gaussian profile and a Moffat
profile were tried. Both fits provided the same relative results,
although the peak was normally fitted more accurately using the
Gaussian profile. No dependence on airmass or time was found, and
median FWHM values and standard errors were found to be: for the
standard target 0\farcs531 $\pm$ 0\farcs002 at 8$^\circ$, 0\farcs566
$\pm$ 0\farcs004 at 98$^\circ$; and for $\eta$ Tel 0\farcs582 $\pm$
0\farcs010 at 8$^\circ$ and 0\farcs545 $\pm$ 0\farcs007 at
98$^\circ$. The lower signal-to-noise on the $\eta$ Tel observations
naturally leads to a  larger degree of scatter in the intensity
profiles.  However it is clear that at Qa there is no 
difference in the FWHM of the $\eta$ Tel frames and standard star
target frames at 98$^\circ$, but that at 8$^\circ$
the FWHM is larger for $\eta$ Tel than for any of
the individual integrations of the standard stars. A further graphical
demonstration of the significance of the extension is provided by
Figure \ref{fig:profiles} which shows the cuts through each Qa 
integration (of 304s) from the standard star imaging (plus signs) and
the $\eta$ Tel observations (dots).  At $8^\circ$ the 
$\eta$ Tel sub-integrations are not only broader at the
FWHM but also have broader wings to their profiles compared to the
standard star images.  Neither of these features is seen at
98$^\circ$. 

The Si-5 image of $\eta$ Tel shown in Figure \ref{fig:cons} includes
only the first 2 of 3 integrations on the source.
The final integration showed an unusual shape compared to the previous
integrations: a greater extension in the North-South direction, and an
unusual asymmetric shape, with an extension towards the East of the image not
matched on the West side.  This shape was seen in all but the
  first nod set.  Neither the North-South extension nor the Easterly
  extension match the chop PA of 55$^\circ$. 
The eccentricities and orientations of the major axis 
for the 3 integrations are:
0.027$\pm$ 0.009 at 110$^\circ$;
0.082 $\pm$ 0.009 at 167$^\circ$;
and 0.163 $\pm$ 0.012 at 80$^\circ$.
Thus we consider that the shape seen in the final integration is not evidence 
for true extension, but is due to the variable seeing conditions on this night.

In the Si-5 residual image we see only low
significance emission at the Northern edge of the original stellar
image. The emission seen in this residual image (Figure \ref{fig:cons}
lower right image) is 6.6 $\pm$ 0.9 mJy (this error includes
  only background error, not PSF uncertainty), however with subtraction of
individual standard star images rather than the coadded standard star
PSF reference image flux in this region can be as low as -7.8mJy,
demonstrating the importance of consideration of the PSF variation
when determining evidence of extended emission close to the star, and
the true insignificance of this emission.  Again the linecuts for
individual nodsets, both for $\eta$ Tel and standard, were
assessed. No dependence on airmass or time was found, and 
median values and standard errors of the Gaussian FWHM measurements
excluding the final integration for $\eta$ Tel (see above) are: for
the standard target 0\farcs383 $\pm$ 
0\farcs003 at 8$^\circ$, 0\farcs407 $\pm$ 0\farcs006 at 98$^\circ$;
and for $\eta$ Tel 0\farcs386 $\pm$ 0\farcs003 at 8$^\circ$ and
0\farcs414 $\pm$ 0\farcs013 at 98$^\circ$.  Thus examination of the
PSF uncertainty indicates that the residual emission 
is likely to be the result of PSF variation over the course of the
observation.


\section{Modelling the emission}

In order to determine the constraints on the radial distribution of emission
provided by the resolved Qa imaging, we considered models for disk
structure of increasing complexity.
For all models it is assumed that the images are comprised of
unresolved stellar flux (of 114 mJy), plus an additional unresolved
disk component (motivated by the hot component inferred by Chen et al. 2006)
of flux $F_{\rm{unres}}$, along with an additional resolved
disk component of flux $F_{\rm{res}}$.
The IRS photometry was used to constrain the total flux so that
$F_{\rm{unres}} + F_{\rm{res}} = $257mJy.
The resolved flux is assumed to arise from an axisymmetric disk of
opening angle 5$^\circ$ (which is not constrained by the modelling)
which lies at an inclination $I$ to our line-of-sight.
Since the emission is consistent with an edge-on disk $I=0$ is assumed
in the modelling, although the extent to which it is edge-on is
constrained later (see section 4.2).
Two types of resolved disk were assumed.
Narrow rings are defined by $r$, the radius of the
mid-point of the disk, in AU;
a finite ring width of $dr/r=0.2$ was assumed, with a constant
surface density.
Extended rings were defined by a continuous distribution of surface
density from some inner radius $r_{\rm{in}}$ to an outer radius
$r_{\rm{out}}$ with surface density following a power-law
$\Sigma \propto r^\gamma$, assuming grains with black body
temperatures and emissivities. 

The total models (disks and point sources) were convolved with the average PSF
(shown in Figure \ref{fig:cons}) for comparison with the observed data. 
The compatibility of each model with the observed structure was tested
by taking line cuts along the direction of extension (at a PA of 8$^\circ$).
The line cuts were averaged over 3 pixels (0\farcs27) in width centred on
the peak of the emission, taken over a length of 50 pixels (i.e.,
$\pm$ 2\farcs25 from the centre) to 
give a 1 dimensional profile for both the
observation of $\eta$ Tel (with error bars) and for each model tested
(e.g., Figure \ref{fig:bestfit}, right).
For each model described below a range of parameters was tried and a $\chi^2$ 
goodness-of-fit test applied to find the best fitting parameters and 
confidence limits (errors from background statistical error).
In addition to the $\chi^2$ from the line-cuts, that from a comparison of
the model images and observation was also considered (e.g., Figure
\ref{fig:bestfit}, left).

To test whether more complicated models are necessary to explain the
images, a Bayesian Information Criterion (BIC) test was applied to
the model fitting: 
\begin{equation}
\label{eq:bic}
  BIC = N \ln (\chi_{\rm{min}}^2) + k \ln N, 
\end{equation} 
where $N$ is the number of data points and $k$ the number of free
parameters, and $\chi_{\rm{min}}^2$ is the minimum $\chi^2$ value found
from each model (\citealt{Schwarz}; see also
  \citealt{liddle,wahhaj2005}). 
This test penalises models for having unnecessary parameters.
A lower value of the BIC is preferred; a difference of 2
between BICs of different models indicates positive evidence against
the higher BIC value, and a difference of 6 indicates strong evidence
against the higher BIC value.  

\subsection{One parameter models}

The simplest model is one with no resolved disk component.
In this case we allowed $F_{\rm{unres}}$ to be a free parameter,
but as expected from the images and comparison of radial
profiles in Figure \ref{fig:profiles}, the $\chi^2$
testing showed that no model provided a good fit to the profile with
$\chi_{\rm{min}}^2 $=786 and a reduced  $\chi_{\rm{r}}^2$=16,
where  $\chi_{\rm{r}}^2$=1.0 is a perfect fit. 
The BIC of this model was 98 (from the 1 dimensional profiles).

Another possibility is to assume that there is no unresolved disk
component, just a narrow ring defined by the free parameter
$r$. Values of $r$ tested in this case were in the range 8--60AU. This
lower limit was chosen as smaller disks appear unresolved and thus
provide the same goodness-of-fit as the unresolved emission only model
described above. 
In this case the fit to the observed profile was poor, worse than assuming the
emission to be completely unresolved with  $\chi_{\rm{min}}^2
$=2815 and a reduced  $\chi_{\rm{r}}^2$=59. 
The BIC of this model was 160.
This demonstrates that the images confirm the suggestion of Chen et al.
that the material is at multiple radii.

\subsection{Two parameter (preferred) model}

\begin{figure*}
\begin{minipage}{8cm}
\includegraphics[width=8cm]{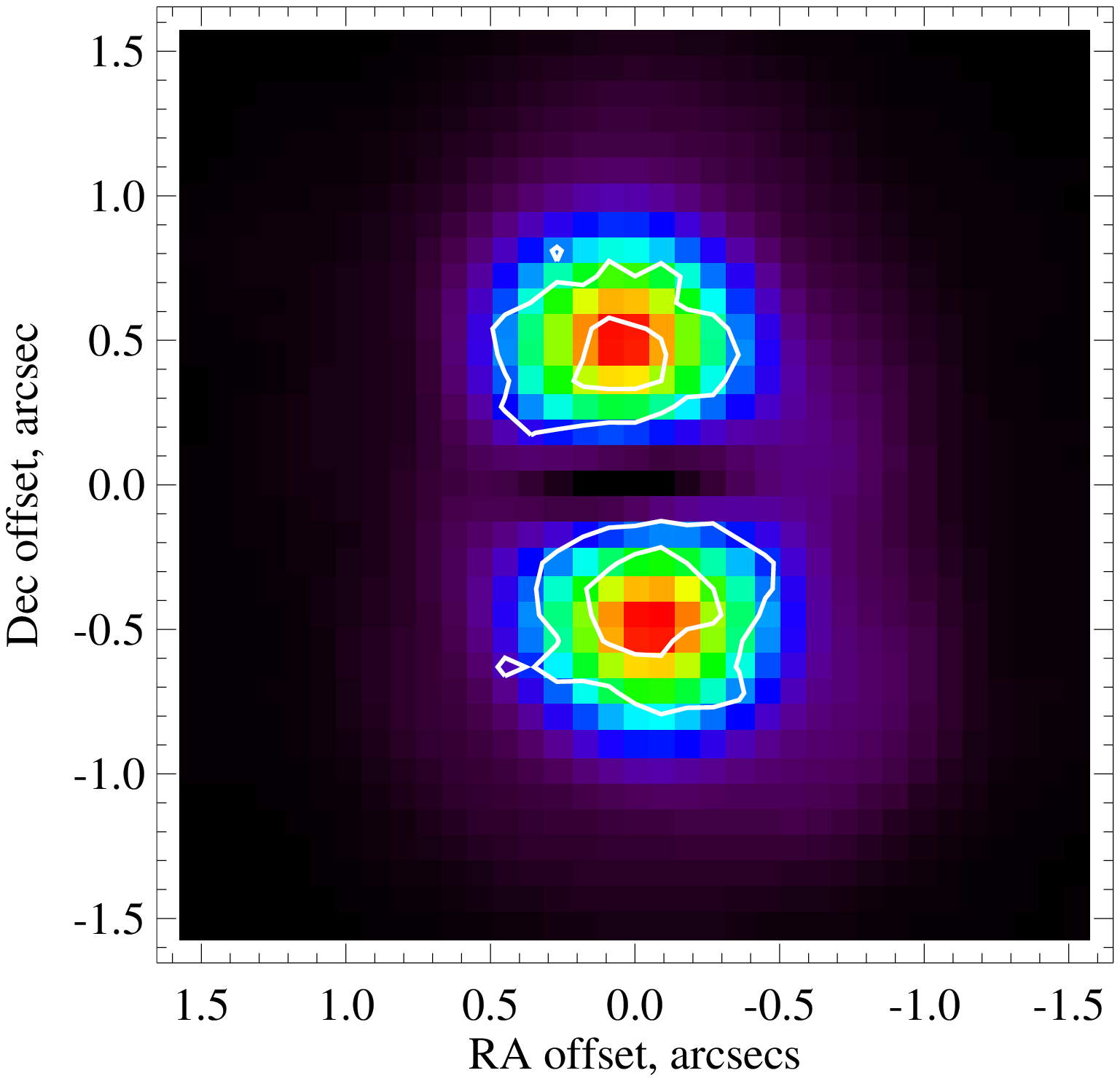}
\end{minipage}
\begin{minipage}{8cm}
\includegraphics[width=8cm]{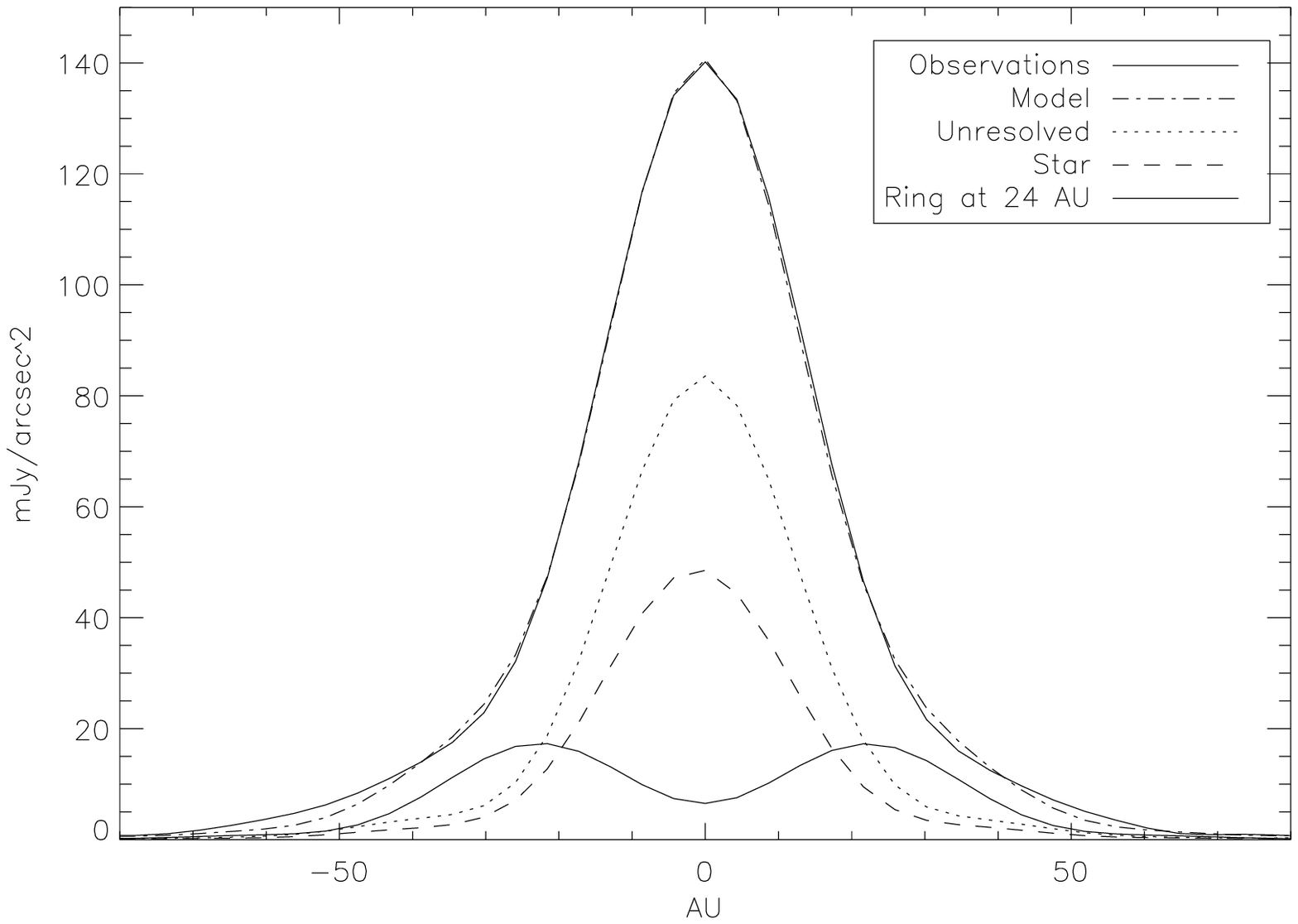}
\end{minipage}
\caption{\label{fig:bestfit} The best fitting two parameter model for
  the Qa band emission around $\eta$ Tel. This model includes that
  stellar flux, an unresolved flux component at 205mJy and and an
  edge-on resolved disk component between 21--26AU. Left: The model
  ``residual'' image, with contours from the observed
    residuals image (Figure \ref{fig:cons} top right).  The model was
  convolved with the PSF and subjected to the same subtraction
  (standard star PSF reference image scaled to peak) as the $\eta$ Tel
  image. The scale is the same as the observed residuals
  image. The observed residuals contours show the good
    correspondence of the model with the observed emission and 
  are at 1/3 and 2/3 of the
  residual peak.   Right: The line cut through the $\eta$ 
  Tel Qa image at 8$^\circ$ and the best fitting model at the same
  position angle. }
\end{figure*}

The two free parameter model assumes that the disk emission is
comprised of a narrow ring of radius $r$ and an unresolved 
component at a level $F_{\rm{unres}}$ (and so is essentially a
  combination of the two one parameter models considered above).
The free parameter $F_{\rm{unres}}$, which also controls the value of 
$F_{\rm{res}}$, was varied between 0 and 280 mJy (in 35 mJy steps),
while $r$ was tested with values between 0 and 95AU (in 3 AU intervals).
A much better fit was achieved with the two parameter model
with an unreduced $\chi_{\rm{min}}^2$ = 53 and a reduced
$\chi_{\rm{rmin}}^2$=1.13 for $r=24$AU and $F_{\rm{unres}}=205$mJy
(corresponding to $F_{\rm{res}}=54$mJy).
The profile of this best fitting model is shown in Figure \ref{fig:bestfit}
in which the contributions of the different components are identified.  
The BIC for this model was 31 (from 1 dimensional profiles)
indicating that this is strongly favoured over the 1 parameter models.

The $\chi^2$ minimisation showed that the parameters were constrained
to the ranges:
$F_{\rm{unres}}=116-225, 136-231, 96-243$ mJy and $r=16-32, 10-42, 10-53$AU
at the 1,2, and $3\sigma$ levels respectively (ranges from a
  marginalisation of the joint distribution of the two parameters).
Using the line cuts perpendicular to the direction of maximum
extension the inclination to our line of sight was constrained to be
less than 20$^\circ$, since at $I\ge20^\circ$ the model FWHM
(0\farcs574) exceeds the observed FWHM (0\farcs566 $\pm$
0\farcs004) by more than 1$\sigma$. 

Given the importance of PSF variation in the residual structure seen
in the N band images, and as we used the average PSF in the modelling outlined
above, we performed the same modelling process using the different PSFs
measured in the single standard star observations (i.e. not co-added).
The sensitivity of our best fit parameters on the chosen PSF was minimal,
with typical 1 $\sigma$ confidence regions close to those given in section 4.2
(largest ranges were 8AU $<$ r $<$ 32AU and 203mJy $<
F_{\rm{unres}} < $ 339mJy). 
Thus the variation in the observed PSF does not strongly change the
best fitting model for the extended emission.

The final total model at Qa is shown after subtraction 
of the PSF scaled to the peak of the observed image of $\eta$ Tel in Figure
\ref{fig:bestfit} along with the profile of this best fitting model
with an edge-on disk (inclination = 0). 
Subtracting the  model image from the observation reveals that 
 two are a  close match within the statistical noise over the area
of the image (900 pixels = 2\farcs7$\times$2\farcs7, reduced $\chi^2$ =
1.3).

\subsection{More complicated models}

As soon as the ring is assumed to be broad the number of free parameters
increases to 3 or 4, depending on whether an unresolved component was
included: i.e., $r_{\rm{in}}$, $r_{\rm{out}}$, $\gamma$, and
possibly $F_{\rm{unres}}$.
We started by assessing whether the radial distribution could be explained by
a single continuous distribution of dust, without an unresolved component.
We tried model grids with $r_{\rm{in}}\in(0.2, 6)$AU spaced at 0.2AU
intervals, $r_{\rm{out}}\in(6, 60)$AU in 1.5AU intervals
and $\gamma$ tested at intervals of 0.5 with $\gamma \in (-2,+2)$.
This resulted in a poor fit to the data ($\chi^2$=198 over the
  1 dimensional profile, reduced
$\chi^2$=4.2), and the BIC of 83 which suggests that the radial
distribution is discontinuous. 

It was possible to find a better fit to the observations than the two
parameter model of section 4.2 when 
$F_{\rm{unres}}$ was also included as a free parameter;
a reduced $\chi^2$ of 1.03 (from the 1d profile) was found for
$r_{\rm{in}} = 1.5^{+2}_{-0.5}$AU, 
$r_{\rm{out}} = 45\pm5$AU,
$\gamma = -0.5$
and $F_{\rm{unres}}=170\pm$26 mJy.
However, the BIC calculation results in a value of 55 suggesting that
the additional parameters required for an extended dust distribution 
are not justified to fit the data.

\subsection{Combining imaging and SED constraints}

\begin{figure*}
\begin{minipage}{10cm}
\includegraphics[width=10cm]{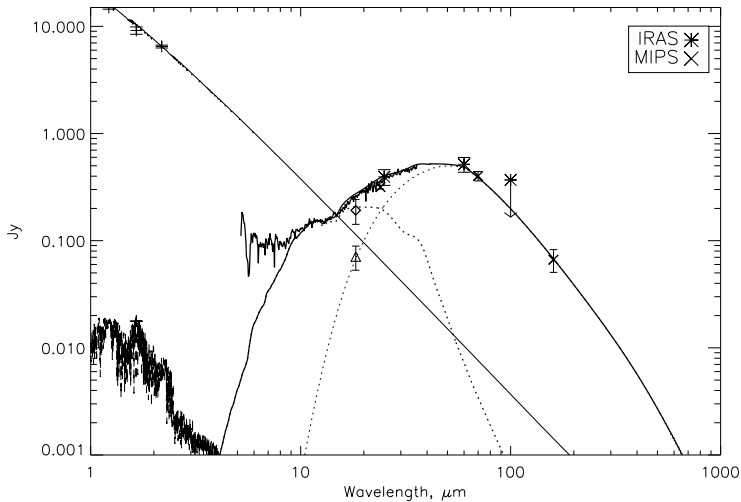}
\end{minipage}
\hspace{0.5cm}
\begin{minipage}{7cm}
\caption{\label{181296sed} Spectral energy distribution (SED)
  of $\eta$ Tel.
  The photosphere of $\eta$ Tel is fitted with a Kurucz model profile scaled
  to the 2MASS K band flux and shown with a solid line, as is the
  photosphere of the binary companion which 
  is modelled with a NextGen model atmosphere (appears on this plot
  only shortwards of 5$\mu$m). 
  Observations at $>5\mu$m are shown after the subtraction of photospheric
  contributions:
  the IRS spectrum of \citet{chen06} (solid line between $\sim$5--30$\mu$m),
  MIPS fluxes (\citealt{rebull}, crosses), IRAS colour-corrected
  excess (asterisks)
  and the contribution to the 18$\mu$m fluxes from the unresolved
  (open triangle) 
  and resolved (open diamond) components, as inferred from the modelling
  of section 4.4 (see also Table \ref{tab:etatelobs}). 
  Note that the IRS excess emission spectrum at
  $\lesssim$ 8$\mu$m is subject to large uncertainty due to
  uncertainties in the level of photospheric emission in this range.
  The two component fit to the excess spectrum is shown by dotted
  lines - the hotter unresolved component at 3.9AU and a resolved component
  at 24AU. }
\end{minipage}
\end{figure*}

The emission spectrum of $\eta$ Tel is shown in Figure
\ref{181296sed}.
Here the stellar photospheric emission has been modelled using a Kurucz
atmosphere of 9506K that has been scaled to the 2MASS K band photometry
(implying a stellar luminosity of 22L$_\odot$ for 48pc).
The photosphere of the M7.5 star companion is shown for reference and
is modelled using 2700K NEXTGEN model atmosphere
\citep{hauschildt}. This companion is outside the aperture
  used to determine the TReCS photometry.  In any case the emission
  from this M7.5 star at the wavelength range covered by both TReCS
  and the Spitzer IRS spectrum is below the background statistical
  noise level and thus it does not contribute to the photometry shown
  at $>$5$\mu$m. 
Photospheric contributions have been subtracted from the fluxes shown
at $>5\mu$m which represent the excess emission from the disk.
The results of our imaging require that the emitting dust be at two
radii, in agreement with the conclusions of \citet{chen06}.  These
separate components are shown by open symbols on Figure
\ref{181296sed}. 

Modelling was performed to provide a two component fit to the SED in which
each component (the unresolved excess and resolved disk) is
pegged to its 18$\mu$m flux, with the unresolved 
component dominating the IRS spectrum $<18\mu$m and the resolved component
dominating the longer wavelength fluxes. We assume a maximum grain
size of 1000$\mu$m, with a minimum grain size of 10.8$\mu$m as this is 
the size at which grains are removed from the system by radiation
pressure. Large grains (greater than mm-sized) are
  inefficient absorbers and emitters of radiation, and their addition
  to the grain population would not have a great affect on the SED. 
A size distribution of $n(D)\propto D^{-3.5}$
\citep[see][]{dohnanyi} was also assumed.  We consider a
  distribution of grain sizes to be a realistic model for a dust
  population created in the different origin scenarios presented in
  section 5 (\citealt{chen06} use a single grain size to fit the IRS
  spectrum). Grains were assumed to have a silicate core and organic
  refractory/icy mantle \citep[as used
    in][]{li&greenberg,augereau99}.  Dielectric constants were
  calculated using Maxwell-Garnett effective medium theory
  and optical properties calculated using Mie theory, Rayleigh-Gans
  theory and Geometric Optics in the appropriate size regimes
  \citep{bohren}. The unresolved component was
then best fit by a silicate fraction of 30\% by volume (with the remaining
volume composed of organic refractory material) and porosity of 0.2,
which fit the shape and level of the silicate feature at
$\sim10\mu$m, resulting in a dust population at 3.9AU 
of fractional luminosity $L_{\rm{dust}}/L_\star = 1.57\times10^{-4}$.
Although we do not consider this modelling to have uniquely constrained
the grain composition, this is a physically based model from which it is
possible to estimate the radial location of the dust.
Since this location is smaller than the FWHM it is confirmed that this should
appear unresolved in our images, which would be expected as long as the radius
was $\leq 6$AU.

\begin{table*}
\centering
\begin{tabular}{*{6}{|c}|} \hline Band & Observed &
  Photospheric & Observed Excess 
 & Expected
  from & Expected from  \\  &  emission, mJy & emission, mJy &
  emission, mJy & unresolved  &
  resolved  \\ & & & & component, mJy & component, mJy \\ \hline
Si-5 & 342 $\pm$ 44 & 282 & 60 $\pm$ 44 & 140 & 6 \\
Qa & 345 $\pm$ 15 & 114 & 231 $\pm$ 15 & 185 & 62 \\ \hline
\end{tabular}
\caption[
Observed excess emission from T-ReCS observations of
  $\eta$ Tel]{\label{tab:etatelobs} The observed emission from the
  T-ReCS observations of $\eta$ Tel, with comparison to the predicted
  emission from the two components used to fit the excess
  SED as described in section 4.4. Recall that the Si5
    photometry suffered from poor calibration accuracy, and as such
    SED modelling was primarily fitted to the IRS spectrum at shorter
    wavelengths. Figure \ref{181296sed} shows the fit recreates
    the IRS spectrum accurately. }
\end{table*}

The same dust grain size distribution was used to model the cold
resolved component.  The same dust composition with the dust located
between 21 and 26AU as constrained by modelling the resolved Qa
emission provides a good fit to the Spitzer IRS spectrum of the excess
emission (Figure \ref{181296sed}). 
The fractional luminosity of the resolved component is similar to the
unresolved component at $L_{\rm{dust}}/L_\star = 1.39\times10^{-4}$.
The N band flux of the resolved component is just 6mJy, which is
consistent with the non-detection of the resolved emission in the
imaging which is dominated by the unresolved component (at 140mJy).
The full SED fit was compared to the observed photometry and a
  $\chi^2$ goodness-of-fit computed.  Data points from IRAS, MIPS,
  TReCS and the IRS spectrum were weighted according to their
  uncertainty.  Data points at $<$7$\mu$m were excluded from the fit
  due to high levels of uncertainty arising from the correct
  determination of the stellar contribution at this range. The
  resulting reduced $chi^2$ of 1.14 indicates the model fits the
  observed emission well.

Thus the two component model provides a consistent picture of this
disk (see Table \ref{tab:etatelobs} and Figure \ref{181296sed} for
comparison between model and observed emission levels),
and while we cannot accurately assess the range of temperatures (and radii)
that are present in the two components, the poor fit of a continuous
surface density distribution (section 4.3) suggests that there is a region
of lower surface density between the two.


\section{Discussion}

In the following section we explore the possible origin of the two
dust populations around $\eta$ Tel.  We consider two possibilities: that
$\eta$ Tel represents an analogue of the Solar System when it was at a
similar age; and that the observed emission could arise from ongoing
planet formation.  We discuss the implications of these scenarios for
any as yet undiscovered planets in the system, and place the $\eta$
Tel disks in a wider context by comparison to the other $\beta$
Pic moving group A star disks. 

\subsection{A young solar system analogue}

The two component model for the $\eta$ Tel disk represents an
intriguing analogue to  the young solar system.
Current theory suggests that the asteroid and Kuiper belts in the Solar
System were originally much more massive than they are today and that much
of the mass depletion occurred during a chaotic period known as the Late
Heavy Bombardment several hundreds of Myr after the formation of the Solar
System, possibly triggered by the migration of the giant planets \citep{gomes}.
According to the \emph{Nice} model \citep{tsiganis},
which reproduces fairly accurately the current configuration of the Solar
Systems giant planets and the Kuiper Belt \citep{levison},
the Kuiper belt was situated between $\sim$15.5--34AU, with 
the orbits of the outer planets confined to the 5-15AU region, whereas
the asteroid belt was at its current location.
Thus in terms of radial location it is tempting to interpret the $\eta$ Tel disk
in terms of an analogous asteroid belt (the unresolved component at 3.9AU)
and an analogous Kuiper belt (the resolved component at 24AU).
It is worth noting that the luminosity of $\eta$ Tel is much
  higher than the Sun at $\sim24L_\odot$, and thus the snow line (the
  location at which ices are expected to form) would be much further
  from the star.  Using a simple 150K blackbody approximation the snow
line would be expected at $\sim$17AU around $\eta$ Tel (compared to
3.5AU in the Solar System).
 
The current levels of excess emission arising from the two
populations of debris in the Solar System are 2-3 orders of magnitude
lower than that observed around $\eta$ Tel, since the fractional
luminosity of asteroid belt dust is
$L_{\rm{dust}}/L_\star \sim$ 10$^{-8}-10^{-7}$ \citep{dermott02astIII},
and that of Kuiper belt dust is $L_{\rm{dust}}/L_\star
\sim$ 10$^{-7}-10^{-6}$ \citep{stern96aa}.
However, the luminosities would have been much larger at an age similar to
$\eta$ Tel, since this would have been before the depletion of the asteroid and 
Kuiper belts.
The original belts in the \emph{Nice} model had masses of
$\sim$1M$_{\oplus}$ and $\sim$35M$_{\oplus}$ respectively
(total mass in full size range of bodies), and so would also
have had correspondingly higher levels of dust emission.
The fractional luminosity of the Kuiper belt was estimated to be
around $5\times 10^{-4}$ during the pre-LHB phase (Wyatt 2008; Booth et
al. in prep.), which is comparable to that of the resolved component of
the $\eta$ Tel disk.

\subsection{Origin of the unresolved component}

The fractional luminosity of the pre-LHB asteroid belt would also have been
considerably higher, since $1M_\oplus$ of material at $\sim 3$AU
distributed in a collisional cascade size distribution from 2000km
asteroids down to 0.8$\mu$m dust blown out by radiation pressure would
have resulted in a fractional luminosity of $\sim 10^{-3}$.
However, collisional processing would ensure that such a high mass would not
be long-lived, and \citet{wyattsmith06} showed how there is a maximum fractional
luminosity for planetesimal belts of a given age and radius.
For a belt at 3.9AU around $\eta$ Tel that has been undergoing
catastrophic disruption for 12Myr this maximum fractional luminosity
is $f_{\rm{max}} = 4 \times 10^{-6}$.
Although the current fractional luminosity is 30-40 times larger than
this maximum  value, this does not rule out the possibility that the
unresolved component arises from the steady state collisional
destruction of an asteroid belt for two reasons. First, the model
requires $f_{\rm{obs}}/f_{\rm{max}} \gg 100$ times larger to indicate
a discrepancy given the various uncertainties in the model parameters
including assumptions about the maximum planetesimal size,
  planetesimal strength and orbital eccentricities of around 0.05 used
to determine $f_{\rm{max}}$
\citep[see also][]{lohne}.
Second, it is possible that collisions in the asteroid belt only became
destructive after the protoplanetary disk dispersed
(after which time eccentricities can increase due to reduced
  dampening), and so more recently than 12Myr thus increasing $f_{\rm{max}}$.

However, another possible interpretation of emission at a few AU is that
it arises in the late stages of the formation of terrestrial planets
\citep[e.g.,][]{rhee, meyer08}.
High levels of hot emission are expected in the planet formation models
of \citet{kenyon04II} once the planetesimal disks are stirred by the
formation of  Pluto-sized objects.
As such objects form within $10^5$ yrs at 4AU in a quiescent disk with the same 
density as the minimum mass solar nebula (MMSN) \citep{kenyon05},
stirring of the disk by a newly formed Pluto does not seem likely to
explain the hotter dust emission.
However, evidence from the Solar System suggests massive collisions between 
terrestrial planet cores could occur quite frequently in the first
$<$100 Myr (e.g.  formation of the Moon, \citealt{canup}), and models
indicate that the final stage of  terrestrial planet formation is
characterised by such collisions \citep[e.g.,][]{Kokubo06}.
Thus the hot emission around $\eta$ Tel could be evidence of
stochastic collisions between large rocky cores.

The unresolved component does not represent a population of
  PR-dragged particles from the resolved ring.  Consideration of the
  collisional and PR-drag timescales for particles from the resolved
  ring \citep[\textbf{$t_{\rm{coll}}/t_{\rm{PR}} \approx 0.05$ for a belt at
    3.9AU;} see][for equations]{wyattins05} indicates that any  
  grains susceptible to PR-drag released by collisions in the outer
  belt are likely to undergo further collisions before PR-drag can
  affect their orbits to move a significant dust population to the
  radius of the unresolved component.  Furthermore, were PR-drag a
  dominant process in this disk system, we would expect a dust
  distribution with a constant surface density \citep{wyattins05}, not
the two separate populations observed in the Qa images. 

It is not possible to distinguish between the asteroid belt and
protoplanet collision  possibilities at this stage, although detailed
spectral modelling may be able to  determine the dust composition and
so give clues as to the nature of the parent body
\citep[e.g.,][]{lisse07}.  Resolving 
the disk could also aid in its interpretation, although the small
spatial scale of this population would make this a challenge.  To date
only the asteroid belt analogue around $\zeta$ Lep has been resolved
at a similar offset from the host star \citep{moerchen}.

\subsection{Origin of the resolved component}

Although the suggested location and fractional luminosity of the resolved
component around $\eta$ Tel compare well with the early Kuiper belt, this
does not require the two systems to have undergone a similar evolution
and to have similar planetary systems.
An alternative scenario for the evolution of the $\eta$ Tel system is given
by the self-stirred planet formation models of \citet{kenyon04II} which
may also explain the delay in the onset of $24\mu$m emission peak until
10-15Myr \citep{Currie}. In such models planet formation within
an extended planetesimal belt results in a bright ring of emission at
the radius where Pluto-sized objects have recently  formed.
The timescale for the formation of Plutos, $600
(\Sigma_0/\Sigma_{\rm{MMSN}})^{-1} r^3$ yr, around $\eta$ Tel suggests
that to have reached this stage at 24AU by 12Myr requires a disk
surface density equivalent to $\sim 0.7$ MMSN. 

To assess the possibility that the bright ring in the $\eta$ Tel disk is 
caused by the recent formation of Pluto-sized objects at 24AU, we performed 
additional modelling of the 18$\mu$m image based on the prescription for
the evolution of dust surface density given in \citet{wyattreview} 
that provides an
empirical fit to the self-stirred models of \citet{kenyon05}.
This model considers an extended planetesimal belt of surface density
scaled to the MMSN in which annuli at different radii have a suppressed dust
content until the age at which Plutos form at which point their mass is eroded
through steady state evolution \citep{wyattsmith06}.
The dust emission is calculated assuming a ratio of dust area to planetesimal
mass appropriate for a collisional cascade size distribution, and further
assuming black body temperatures and emission properties.
To account for possible $\textbf{global}$ 
deviations from this size distribution or from
pure black body grains, an additional scaling factor $\eta$
was applied to the surface brightness of the whole disk. An unresolved
disk component was included in addition to the unresolved star.  The
total flux from the self-stirred disc (as fixed by the disk surface
density and scaling factor) and the unresolved component was fixed to
259 mJy from the IRS photometry (see section 4). 
Thus the model has 2 free parameters: $\Sigma_0/\Sigma_{\rm{MMSN}}$,
and $\eta$, with $F_{\rm{unres}}$ fixed in response to the level of
flux in the self-stirred disk. 

\begin{figure}
\begin{minipage}{8cm}
\includegraphics[width=8cm]{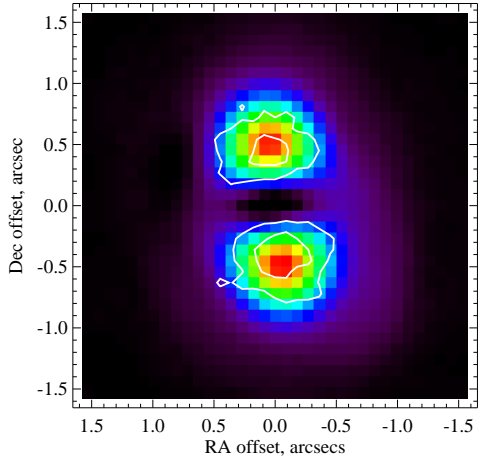}
\end{minipage}
\caption{\label{delstirres} The self stirred model of the resolved
  disk component as described in Section 5.3 after the same PSF
  subtraction as applied to the observations (Figure \ref{fig:cons}
  Qa band residual image) and 2 parameter model,
  shown in Figure \ref{fig:bestfit}.  The scaling in this image is the
same as for Figure \ref{fig:cons} top left. Contours are from
  the observed residuals image, at levels of 1/3 and 2/3 of the peak,
  as for Figure \ref{fig:bestfit}. }
\end{figure}

As expected from the simple calculation above, the best fitting model has
a surface density of 0.7 times MMSN, leading to a ring at $\sim 24$AU,
with $F_{\rm{unres}}=216$mJy (similar to the model of section 4.2) and
$\eta=0.22$.
The low value of $\eta$ could arise because the grains are cooler than
black body, since the ratio of $B_\nu(105K)/B_\nu(123K)$ is $\sim 0.3$
at 18$\mu$m.
The $\chi^2$ for this model is 58 with a reduced $\chi^2$ of
$\chi_{\rm{min}}^2$=1.21, slightly higher than that of the 2 parameter
model ($\chi_{\rm{min}}^2$=1.13) in Section 4.2. The reduced $\chi^2$
of the image (taken as described in Section 4.2) was 1.15, better than
the two parameter model which had $\chi_{\rm{min}}^2$=1.3. The BIC for
this model is 32, which is indicative of the almost equally good fit
of the two models. The BIC values show there is no evidence against
either the simple ring plus unresolved flux model (section 4.2) or the
self-stirred model.  The residuals for the best fitting self-stirred
model are plotted in Figure \ref{delstirres}.  Multi-wavelength
imaging may help break this degeneracy.
The self-stirred model requires an extended disk with a brightness
peak at the location of recent Pluto formation.  If a larger disk
location were observed in longer wavelength imaging this could be
evidence in favour of the self-stirred interpretation, whereas in the
solar system analogy we may expect to see the disk confined to a
narrow radial range. Alternatively searching for planets in the system
could further test these models. 

In other words the structure of the disk is entirely consistent with
a self-stirred extended planetesimal belt in which the inner region
has been cleared through collisional erosion following the formation
of Pluto-sized objects.
One caveat is that an additional unresolved component still has to be
invoked.
To be consistent such a component could not originate in an asteroid
belt (unless this had significantly different material properties to
the outer regions), and further work is needed to assess the duration
and radial location of any 18$\mu$m emission expected from the final 
collisional phase of terrestrial planet formation.

\subsection{Implications for the planetary system}

Taking the solar system analogy to its extreme would suggest that the
gap in the dust distribution seen around $\eta$ Tel at 5-20AU is caused
by the existence of giant planets distributed in this region.
Like the Solar System, the planetesimal belt at 3.9AU could represent a region
where terrestrial planet formation was disrupted by the influence of a
gas giant planet at 5-6AU, and the inner edge of the belt at 24AU could demark
the outer edge of the planetary system.  

The inner edge of the the hot dust could also
be evidence of terrestrial planets at $<$3AU, as the hot dust at 1AU around
HD\,69830 \citep{beichman05, lisse07} lies just outside 3
Neptune mass planets, the outermost of which is at 0.63 AU
\citep{lovis} (although the location of the disk has not yet
  been confirmed through resolved imaging).  Also, although no
planets have been detected around 
zeta Lep, the restricted spatial extent of its asteroid belt
\citep{moerchen} is suggested to be possible evidence of perturbations
of planetary bodies.

Implicit in the self-stirring interpretation is the formation of
planets at least the size of Pluto in the inner regions.
Such objects would presumably have grown through mutual collisions to
larger sizes, but it is not required that they also evolved into gas
giant planets, which would depend on whether they reached $\sim
10M_\oplus$ before the gas disk dissipated \citep[e.g.][]{lissauer_rev}.

A further possibility which is not discussed here is that this disk
is a remnant of the protoplanetary disk, and that the dust is shepherded
into a ring through the action of gas drag forces or instabilities
\citep{takeuchi, klahr, besla}. In this case it would not be necessary
for any planets to be present.  A massive gas disk around a
  star of this age would be highly unusual, as most gas disks are
  believed to dissipate in $<$ 10Myr \citep[e.g.][]{hollenbach05},
  although gas has been detected around 4 A stars ($\beta$ Pic
  \citealt{brandeker}; HD\,141596 \citealt{jonkheid}; 49 Ceti
  \citealt{dent}; and HD\,32297 \citealt{redfield}). 
To assess this possibility knowledge of the gas distribution is vitally
important, yet gas has yet to be detected around $\eta$ Tel. 

\subsection{Comparison with A star disks in $\beta$ Pic moving group}

Of the 5 A stars in the $\beta$ Pic moving group, 3 exhibit excess mid-IR
emission from a debris disk ($\beta$ Pic, $\eta$ Tel and HD\,172555).
The derived (or directly measured) structures of these 3 disks display
significant diversity.
$\beta$ Pic, an A6V star, has a well known highly extended disk
with dust imaged from 10s of AU to an outer extent of $\sim$1500AU
\citep{telesco, larwood}.
Since dust at the outer edge is thought to be created in planetesimal belts
much closer to the star \citep{augereau},
it is probably most instructive to consider this disk as an extended
planetesimal belt out to ~100AU with a broad peak in
surface density at around 70AU \citep{telesco}. 
HD\,172555 on the other hand has all of its emission much closer to the star.
Its emission spectrum indicates the presence of silicate grains at 520K and a 
blackbody component at 200K implying a radial offset of 0.8--6AU
\citep{chen06, rebull}. 
The $\sim$24AU outer belt of the $\eta$ Tel debris lies intermediate between 
these two extremes.
These disks thus represent the diversity that is seen amongst debris disks at
the age when emission is seen to peak \citep{Currie}. 

Within the context of the self-stirring model this diversity would be expected 
to arise primarily from different initial disk masses.
While $\eta$ Tel is the product of a 0.7x MMSN disk, the protoplanetary disk of 
HD\,172555 (and of the two A stars which are not detected) is presumably of lower
density, and that of $\beta$ Pic is of higher density to allow Plutos
to form out  to 70AU by 12Myr. It is also worth noting that the beta Pic
disk shows an asymmetry that may be the result of a catastrophic collision
at $\sim$50AU, such as might be expected from the self-stirring models at the
epoch of Pluto-formation.
Such a diversity of protoplanetary disk masses is expected
\citep[e.g.,][]{andrews}, but the small number of stars in this sample
does not permit any definite conclusions to be drawn.
In the young Solar System interpretation, the diversity in these disks
could instead be due to the diversity in planetary system architecture
across the three stars, which in addition to a dependence on initial
disk mass \citep{wyatt07} could also have a significant stochastic component
\citep[e.g.][find small changes in the location of planetary embryos
has a great impact on the final system architecture]{quintana}.
Furthermore the origin of the hot emission, and its diversity, remains
a puzzle, and it is not clear whether the presence, or lack thereof,
of asteroid belts and ongoing  terrestrial planet formation should be
a monotonic function of the initial disk mass \citep[see e.g.][who
find from a sample of middle-aged disks around A0--A3V stars that
factors beyond variation in initial disk mass are needed to explain
the variation in observed disk structure]{su08}.


\section{Conclusions}

We have presented Si-5 and Qa band Gemini T-ReCS imaging of the 12Myr old A
star $\eta$ Tel. The emission at 18.3$\mu$m is shown to be
significantly extended compared with a point source.  This
  image represents the first resolution of dust emission around this
  star.  
These observations confirm the interpretation of the SED as a debris disk
comprised of two populations:
one component that is resolved at 24AU,
the other that is unresolved with a temperature consistent with dust at 3.9AU.

Two interpretations are proposed for the origin of the architecture
of these debris belts.
\textbf{(i)} The system is a close analogue to the Solar System at a
similar age, both 
in terms of the radial location of the debris and the level of emission;
i.e., these belts represent analogous asteroid and Kuiper belts.
\textbf{ii)} Alternatively this system is still undergoing planet
formation, and the  
24AU ring is the location where Pluto-sized objects recently formed;
the 3.9AU ring could be the product of recent collisions between Mars-sized
bodies in the final accumulation phases of terrestrial planets.
This implies that this system evolved from a 0.7x MMSN protoplanetary disk.

These possibilities can be further tested by probing the radial structure of
the disk through multiple wavelength imaging, and by searching for planets
both through direct methods, or through looking for evidence for them in the
disk structure.
Thus $\eta$ Tel is a valuable source for testing planet formation theories, 
not least since it may be an example of a system where planet formation is
ongoing.

\begin{acknowledgements}
  RS is grateful for the support of a Fellowship of the Royal
  Commission for the Exhibition of 1851.  LJC is grateful for the
  support of an STFC studentship.  MCW is grateful for the support of
  a Royal Society University Research Fellowship. MM acknowledges
  support from JPL funded by NASA through the Michelson Fellowship
  Program. JPL is managed for NASA by the California Institute of
  Technology.  The authors would like to thank Christine Chen for
  providing the Spitzer IRS spectrum of $\eta$ Tel. 

Based on observations obtained at the Gemini Observatory, which is
operated by the Association of Universities for Research in Astronomy,
Inc., under a cooperative agreement with the NSF on behalf of the
Gemini partnership: the National Science Foundation (United States),
the Science and Technology Facilities Council (United Kingdom), the
National Research Council (Canada), CONICYT (Chile), the Australian
Research Council (Australia), Ministério da Ciência e Tecnologia
(Brazil) and SECYT (Argentina). 

\end{acknowledgements}

\bibliographystyle{aa}  
\bibliography{/data/cass65a/rsed/rs/THESIS/thesis} 

\end{document}